\documentclass[10pt, conference]{IEEEtran}
\usepackage{algorithm}
\usepackage{algorithmic}

\ifCLASSINFOpdf
   \usepackage[pdftex]{graphicx}
\else
   \usepackage[dvipdfmx]{graphicx}
\fi
\usepackage[cmex10]{amsmath}
\usepackage{amssymb}

\usepackage{url}

\usepackage{bm}
\DeclareBoldMathCommand\boldlangle{\left\langle}
\DeclareBoldMathCommand\boldrangle{\right\rangle}

\newcommand{\trigram}[3]{$\langle ~ \texttt{#1},~ \texttt{#2},~ \texttt{#3} ~ \rangle$}
\newcommand{\boldtrigram}[3]{$\boldlangle ~ \texttt{\textbf{#1}},~ \texttt{\textbf{#2}},~ \texttt{\textbf{#3}} ~ \boldrangle$}

\begin{document}
%
\title{Source File Set Search for \\ Clone-and-Own Reuse Analysis}


\author{\IEEEauthorblockN{Takashi Ishio\IEEEauthorrefmark{1}\IEEEauthorrefmark{2}, Yusuke Sakaguchi\IEEEauthorrefmark{1}, Kaoru Ito\IEEEauthorrefmark{1}, Katsuro Inoue\IEEEauthorrefmark{1}}
\IEEEauthorblockA{\IEEEauthorrefmark{1} Graduate School of Information Science and Technology,
Osaka University, Osaka, Japan}
\IEEEauthorblockA{\IEEEauthorrefmark{2} Graduate School of Information Science,
Nara Institute of Science and Technology, Nara, Japan \\
Email: \{ishio, s-yusuke, ito-k, inoue\}@ist.osaka-u.ac.jp }
}


%


\maketitle

\begin{abstract}
Clone-and-own approach is a natural way of source code reuse for software
developers.
To assess how known bugs and security vulnerabilities of a cloned component affect an application, developers and security analysts need
to identify an original version of the component and understand how the cloned component is different from the original one.
Although developers may record the original version information in a version
control system and/or directory names, such information is often either
unavailable or incomplete.
In this research, we propose a code search method that takes as input a set of
source files and extracts all the components including similar files from a software ecosystem (i.e., a collection of existing versions of software packages).
Our method employs an efficient file similarity computation using b-bit minwise hashing technique.
We use an aggregated file similarity for ranking components.
To evaluate the effectiveness of this tool, we analyzed 75 cloned components in Firefox and Android source code.
The tool took about two hours to report the original components from 10 million files in Debian GNU/Linux packages.
Recall of the top-five components in the extracted lists is 0.907, while recall of a baseline using SHA-1 file hash is 0.773, according to the ground truth recorded in the source code repositories.
\end{abstract}


%
\IEEEpeerreviewmaketitle

\section{Introduction}
\label{sec:introduction}

Software developers often reuse source code of existing products to develop a new software product~\cite{Dubinsky2013}. 
Mohagheghi et~al. reported that reused components are more reliable than non-reused code~\cite{MohagheghiICSE2004}.
While open source software projects reuse code from other OSS projects, industrial developers also use open source systems due to their reliability and cost benefits~\cite{10.1109/MS.2008.67}.

Clone-and-own approach is one of the popular approaches to source code reuse~\cite{RubinSPLC2012,Mende2009}.
Dubinsky et~al.~\cite{Dubinsky2013} reported that cloning is perceived as a natural reuse approach by the majority of practitioners in the industry. 
Although many reusable components are available online in binary forms for various operating systems, 
developers copy source code of an existing component into their project's so that they can build and test their product using a particular version of a component or modify it for their own purpose.
For example, \textsl{Mozilla Firefox 45.0} includes a modified version of \textsl{zlib 1.2.8} in its \verb|modules/zlib| directory; a developer added a header file \verb|mozzconf.h| in order to rename functions defined in the library.
Koschke et~al.~\cite{KoschkeIWSC2016} reported that copies of specific libraries are involved in a relatively large number of projects.

Cloned components may introduce potential defects into an application.
Sonatype reported that many applications include severe or critical flaws inherited from their components~\cite{ITWorld2015}. 
Hemel et~al.~\cite{Hemel2012} reported that each of Linux variants embedded in electronic devices has its own bug fixes.

To investigate known bugs and security vulnerabilities of a cloned component, developers and security analysts need to identify an original version of the component and understand how the cloned component is different from the original one.
However, in general identifying the original version is tedious and time-consuming.
The main reason is that original component names and version numbers are often unrecorded~\cite{XiaJSSST2013}.
Another reason is that a cloned component may be a derived version in a different project.
For example, Firefox includes another copy of zlib in \verb|security/nss/lib/zlib| directory; the version is a part of the NSS component.
To identify an original version of a cloned component, an analyst has to compare its source code with all the existing versions of components in its software ecosystem.

A baseline method to analyze a cloned component is code comparison using file hash values such as SHA-1 and MD5.
Since the method cannot detect modified files, Kawamitsu et~al.~\cite{KawamitsuSCAM2014} proposed a code comparison technique that identifies the most similar file revision in a repository as an original version of a file.
The experiment reported 20\% of cloned files are modified in eight projects. 
Although the method is effective, its simple pairwise comparison of files is inefficient to analyze an entire software ecosystem, which includes millions of source files.


In this research, we propose a code search method tailored for analyzing a cloned component.
It takes as input a set of source files, and reports existing components including files that are similar to the input files.
The method ranks components using aggregated file similarity, assuming a cloned component is the most similar to the original component.
A reported list of components enables developers and security analysts to compare their cloned component with its original version.

Our code search method employs the $b$-bit minwise hashing technique~\cite{b-bit} that is an extension of Min-Hash technique~\cite{Leskovec2011,Broder1997};
in summary, the technique enables to estimate file similarity using hash signatures.
Our method constructs a database of hash values for each file in a software ecosystem.
Using the database, our method then extracts a subset of files likely similar to a query and then computes actual similarity for the subset.
Although a database construction takes time, we can search similar files within a practical time.

We also define a filtering method for components using file similarity.
Since different versions of a library often include similar files, 
we select components having the most similar files as a representative set.

We conducted an experiment to evaluate the effectiveness of the tool.
As the ground truth, we manually identified original versions of 75 cloned components included in source code of \textsl{Firefox 45.0} and \textsl{Android 4.4.2\_rc1}.  
We then analyzed the components with a database of Debian GNU/Linux packages including 10 million source files.
Recall of the top-five components in the extracted lists is 0.907, while recall of a baseline using SHA-1 file hash is 0.773.
To obtain all the original components, a user needs to investigate 551 components in the lists.
It is smaller than 931 components reported by the baseline method.
The result shows that our method ranks the original components at higher positions and reduces manual effort of a user.

The contributions of the paper are summarized as follows.
\begin{itemize}
\item We defined a code search method to extract similar files from a huge amount of source files efficiently.

\item We defined a component filtering method to select likely original components.

\item We created the ground truth dataset of actual clone-and-own reuse instances for two major OSS projects, and evaluated our method using the dataset. 

\end{itemize}


Section~\ref{sec:background} shows related work of our approach. 
The approach itself is detailed in Section~\ref{sec:approach}. 
Section~\ref{sec:evaluation} presents the evaluation of our approach using OSS projects.
Section~\ref{sec:threats} describes the threats to validity of the work.
Section~\ref{sec:conclusion} describes the conclusion and future work.


\section{Related Work}
\label{sec:background}

\subsection{Code Clone Detection}

Code clone detection has been used to analyze source code reuse between projects.
Kamiya et~al.~\cite{KamiyaTSE2002} proposed CCFinder to detect similar code fragments between files.
German et~al.~\cite{GermanMSR2009} used CCFinder to detect code siblings reused across FreeBSD, OpenBSD and Linux kernels.
They identify the original project of a code sibling by investigating the source code repositories of the projects.
Hemel et~al.~\cite{Hemel2012} analyzed vendor-specific versions of Linux kernel using their own clone detection tool. 
Their analysis showed that each vendor created a variant of Linux kernel and customized many files in the variant.

Koschke et~al.~\cite{KoschkeIWSC2016} also used a clone detection technique to analyze code clone rates in 7,800 OSS projects.
They found that a relatively large number of projects included copies of libraries.
They excluded the copies from analysis, because the analysis did not focus on inter-project code reuse.

Krinke et~al.~\cite{KrinkeIWSC2010} proposed to distinguish copies from originals using the version information recorded in source code repositories. 
Krinke et~al.~\cite{KrinkeMSR2010} used the approach to analyze GNOME Desktop Suite projects.
The result shows that there is a lot of code reuse between the projects.
Although the version information is useful to select older files as candidates of reused files, our method does not use it because source code repositories are not always available.

Sajnani et~al.~\cite{SajnaniICSE2016} proposed SourcererCC, a scalable code clone detection tool.
They optimized comparison of two code fragments, based on an observation that most of files are different from one another.
We employ $b$-bit minwise hashing technique to avoid unnecessary code comparison.  Our approach can be combined with SourcererCC's optimization.

Sasaki et~al.~\cite{SasakiMSR2010} proposed a file clone detection tool named FCFinder.
The tool normalizes source files by removing code comments and white space, and compare the resultant files using MD5 hash.
This method is not directly applicable to our problem, because it cannot detect similar but modified files.

Hummel et~al.~\cite{HummelICSM2010} proposed to use an index database for instant code clone detection.
While it is similar to our approach, the clone index is designed to report source code locations.
It is not suitable to compute source file similarity.

Jiang et~al.~\cite{Jiang2007} proposed DECKARD, a code clone detection tool using a vector representation of an abstract syntax tree of source code. 
Nguyen et~al.~\cite{NguyenFASE2009} extended a vector representation for a dependence graph and proposed a code clone detection tool named Exas.
The tool has been used to detect common patterns in source code~\cite{NguyenMSR2016}, rather than identification of similar files.

Detected instances of source code reuse are clues to extract the common functionalities in software products.
Rubin et~al.~\cite{RubinSPLC2013} reported that industrial developers extract reusable components as core assets from existing software products.
Bauer et~al.~\cite{Bauer2013} proposed to extract code clones across products as a candidate of a new library.
Ishihara et~al.~\cite{IshiharaWCRE2013} proposed a function-level clone detection to identify common functions in a number of projects.
Our method can be seen as a file-level detection of cloned components.

While our method enables a user to select original components for comparison, it does not directly support a source code comparison activity itself.
Duszynski~\cite{DuszynskiWCRE2011} proposed a code comparison tool to analyze source code commonalities from a number of similar product variants.
Sakaguchi et~al.~\cite{SakaguchiVISSOFT2015} also proposed a code comparison tool that visualizes a unified directory tree for source files of several products.
Fischer et~al.~\cite{FischerICSME2014} proposed to extract common components from existing product variants and compose a new product.

\subsection{Origin Analysis}

Godfrey et~al.~\cite{Godfrey2005} proposed origin analysis to identify merged and split functions between two versions of source code.
The method compares identifiers used in functions to identify original functions.
Steidl et~al.~\cite{SteidlMSR2014} proposed to detect source code move, copy, and merge in a source code repository.
The method identifies a similar file in a repository as a candidate of an original version. 

Kawamitsu et~al.~\cite{KawamitsuSCAM2014} proposed to identify an original version of source code in a library's source code repository.
It is an extension of origin analysis across two source code repositories.
A user of the method must know what library is included in the program.  
Our method does not need such knowledge, because it compares input files with all the existing components in a software ecosystem.

Sojer et~al.~\cite{SojerCACM2011} pointed out that ad-hoc code reuse from the Internet has a risk of a license violation. 
Inoue et~al.~\cite{InoueICSE2012} proposed a tool named Ichi-tracker to identify the origin of ad-hoc reuse.  
It is a meta-search engine to obtain similar source files on the Internet and visualizes the similarities.  
While Ichi-tracker takes a single file as a query, 
our method enables a user to analyze a set of files as a component. 

Kanda et~al.~\cite{Kanda2013} proposed a method to recover an evolution history of a product and its variants from their source code archives without a version control.
The approach also compares the full contents of source files, using a heuristic that developers tend to enhance a derived version and do not often remove code from the derived version.
It might complement our approach, because it helps to understand an evolution history of components reported by our method.

Antoniol et~al.~\cite{Antoniol2000} proposed a method to recover the traceability links between design documents and source files.  
The method computes similarity of classes by aggregating similarity of their attribute names to identify an original class definition in design documents.
Our method can be seen as a coarse-grained extension that identifies original components using aggregated file similarity.

Hemel et~al.~\cite{HemelMSR2011} proposed a binary code clone detection to identify code reuse violating software license of a component.
The method compares the contents of binary files between a target program and each of existing components.
S{\ae}bj{\o}rnsen et~al.~\cite{SaebjornsenISSTA2009} proposed a clone detection for binary code.
It uses a locality sensitive hashing to extract similar code fragments in binary files.
Qiu et~al.~\cite{QiuSANER2015} proposed a code comparison method for a binary form to identify library functions included in an executable file.

For Java software, Davies et~al.~\cite{DaviesMSR2011,DaviesESE2013} proposed a file signature to identify the origin of a jar file using classes and their methods in the file ignoring the details of code. 
German et~al.~\cite{GermanIEEESoftware2012} demonstrated the approach can detect OSS jar files included in proprietary applications.
Mojica et~al.~\cite{MojicaIEEESoftware2014} used the same approach to analyze code reuse among Android applications.
Ishio et~al.~\cite{IshioMSR2016} extended the analysis to automatically identify libraries copied in a product.
Differently from those approaches, our method directly compares source files because small changes in files might be important to understand differences between a cloned component and its original version.

Luo et~al.~\cite{LuoFSE2014} proposed a code plagiarism detection applicable to obfuscated code.
The detection method identifies semantically equivalent basic blocks in two functions.
Chen et~al.~\cite{ChenICSE2014} proposed a technique to detect clones of Android applications.
The analysis uses similarity between control-flow graphs of methods.
Obfuscated code is out of scope of our method.
Ragkhitwetsagul et~al.~\cite{RagkhitwetsagulSCAM2016} evaluated the performance of code clone detection and relevant techniques for source files modified by code obfuscators and optimizations.


\section{Our Approach}
\label{sec:approach}

\begin{figure}[t]
\centering
\includegraphics[width=0.8\linewidth]{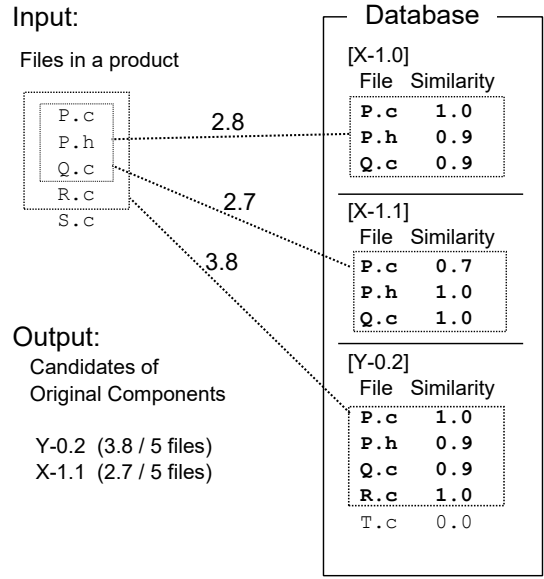}
\caption{An example input and output of our method.  Two components are selected to cover four files.  They are sorted by aggregated file similarity.}
\label{fig:concept}
\end{figure}

Our method takes as input a set of source files and reports a list of components that likely include the original version of the files in a software ecosystem.
In this paper, a software ecosystem is a collection of components $\{ C_1, C_2, \cdots, C_n \}$. 
Each component comprises a set of files.   
Our implementation assumes that each component has a unique name such as ``zlib-1.2.8'' and ``libpng-1.6.9''.  

We use source file similarity, because popular libraries written in C/C++ are used by many projects.
While the main target of our method is C/C++, our implementation supports C/C++ and Java. 
Our method is language independent except for the lexical analysis step.
The lexical analysis assumes that each source file has a correct file extension representing a programming language.

Our method comprises two steps: component search and component-ranking.
The first step extracts a set of components $R$ including files similar to query files $Q$.
The second step filters and ranks components according to aggregated file similarity.
We use a simple assumption: A component in a database is likely original if it has the most similar file to a query file.

Fig.~\ref{fig:concept} shows an example input and output of our method.
The example query $Q$ includes five files: \texttt{P.c}, \texttt{P.h}, \texttt{Q.c}, \texttt{R.c}, and \texttt{S.c}.
Our method compares each query file with files in a component database.
In the example, our component search step detects similar files in three components \texttt{X-1.0}, \texttt{X-1.1}, and \texttt{Y-0.2}.  
Three edges connecting a product and components have labels indicating the sum of similarity values.
Using file similarity and the aggregated similarity, our method ranks \texttt{Y-0.2} at the top and \texttt{X-1.1} at the second.
\texttt{Y-0.2} is the most likely original component, because it has four similar files.
\texttt{X-1.1} is the second, because its two files \texttt{P.h} and \texttt{Q.c} are more similar than files in \texttt{Y-0.2}.  Those files are also likely original files.
Our method filters out \texttt{X-1.0}, 
because it does not have a file whose similarity is higher than other components.

\subsection{Component Search}

Our component search uses a file-by-file comparison.
Given a query set of source files $Q$, we extract candidates of original components $R$ from a collection of components as follows.
\begin{eqnarray*}
R = \{ C_i ~ | ~ \exists q \in Q, f \in C_i. ~ sim(q, f) \geq th \}
\end{eqnarray*}
where $sim$ is a similarity function and $th$ is a similarity threshold, respectively. 
It should be noted that the definition does not use release date of components,
because there is a time lag between an official release of a project and its packaging for a software ecosystem.
If accurate timestamps are available for both query files and components, our method can use a subset of components older than query files.

Our similarity of source files is Jaccard index of token trigrams defined as follows.
\begin{eqnarray*}
sim(f_1, f_2) = \frac{|trigrams(f_1) \cap trigrams(f_2)|}{|trigrams(f_1) \cup trigrams(f_2)|}
\end{eqnarray*}
where $trigrams(f)$ is a multiset of trigrams extracted from a file $f$.
We employ the Jaccard index because it approximates the edit distance~\cite{Ukkonen1992}.
A higher similarity indicates that a larger amount of source code could be reused.
Compared with the longest common subsequence, it is less affected by moved code in a file.
Furthermore, it can be efficiently estimated using the Min-Hash technique~\cite{Leskovec2011,Broder1997}.

\begin{figure}[t]
\centering
{\small
\begin{verbatim}

 Example code:
   a: while ((*dst++ = *src++) != '\0');
   b: while (*dst++ = *src++);
\end{verbatim}
}
{\scriptsize
\renewcommand{\arraystretch}{1.5}
\begin{tabular}{l | l} \hline
$trigrams(a)$ & $trigrams(b)$ \\ \hline
\trigram{\_}{\_}{while}, \trigram{\_}{while}{(}, &  \trigram{\_}{\_}{while}, \trigram{\_}{while}{(}, \\
\boldtrigram{while}{(}{(}, \boldtrigram{(}{(}{*}, & \boldtrigram{while}{(}{*}, \\
\trigram{(}{*}{dst}, \trigram{*}{dst}{++}, & \trigram{(}{*}{dst}, \trigram{*}{dst}{++}, \\
\trigram{dst}{++}{=}, \trigram{++}{=}{*},  & \trigram{dst}{++}{=}, \trigram{++}{=}{*}, \\
\trigram{=}{*}{src}, \trigram{*}{src}{++}, & \trigram{=}{*}{src}, \trigram{*}{src}{++}, \\
\trigram{src}{++}{)}, & \trigram{src}{++}{)}, \\
\boldtrigram{++}{)}{!=}, \boldtrigram{)}{!=}{'$\backslash$0'}, & \boldtrigram{++}{)}{;}, \\
\boldtrigram{!=}{'$\backslash$0'}{)}, \boldtrigram{'$\backslash$0'}{)}{;}, & \\ 
\trigram{)}{;}{\_}, \trigram{;}{\_}{\_}   & \trigram{)}{;}{\_}, \trigram{;}{\_}{\_}  \\ \hline
\end{tabular}
}
{\small
\begin{eqnarray*}
sim(a, b) =  \frac{|trigrams(a) \cap trigrams(b)|}{|trigrams(a) \cup trigrams(b)|} = \frac{11}{19} = 0.579 \\
\end{eqnarray*}
}
\caption{An example of similarity value.  A bold trigram is unique to a code fragment.}
\label{fig:trigrams}
\end{figure}

We use a token as a trigram element to ignore the length of identifier names.
A lexer extracts a token sequence by removing comments and white space.
The lexer keeps identifiers as they are, because identifiers are important clues to identify a version~\cite{KawamitsuSCAM2014}.
Our lexer also keeps preprocessor directives in C/C++ source files.
Fig.~\ref{fig:trigrams} shows a pair of example code fragments, their trigrams, and a similarity value obtained from the trigrams.
In the figure, a trigram \trigram{A}{B}{C} indicates three consecutive tokens in a code fragment.
A symbol ``\_'' in a trigram indicates the beginning and end of a file.

\renewcommand{\algorithmicinputs}{\textbf{Inputs}}
\renewcommand{\algorithmicoutputs}{\textbf{Outputs}}

\begin{algorithm}[t]
\caption{Component Search}
\label{fig:algorithm}

\begin{algorithmic}
\INPUTS
\item $Q$: A set of files to be analyzed.
\item $F$: A set of existing files in a software ecosystem.  
\item $Owners(f)$: A mapping from a file $f$ to components including the file.
\ENDINPUTS

\OUTPUTS
 \item $R$: A set of components including query files.
 \item $S(q, C)$: Similarity of a query file $q$ and its most similar file in a component $C$.
\ENDOUTPUTS
\end{algorithmic}
\begin{algorithmic}[1]

  \STATE{Initialize $R \leftarrow \phi$}
  \STATE{Initialize $S(q, C) \leftarrow 0$ for all possible $q$ and $C$.}
  \FOR{$q \in Q$}
    \FOR{$f \in F$}
      \IF{$\frac{\min(|trigrams(q)|, |trigrams(f)|)}{\max(|trigrams(q)|, |trigrams(f)|)} \geq th$}
        \IF{$sim_e(q, f) \geq th-m$}
          \IF{$sim(q, f) \geq th$}
            \FOR{$C_i \in Owners(f)$}
              \STATE{$R \leftarrow R \cup C_i$}
              \IF{$S(q, C_i) < sim(q, f)$}
                \STATE{$S(q, C_i) \leftarrow sim(q, f)$}
              \ENDIF
            \ENDFOR
          \ENDIF
        \ENDIF
      \ENDIF
    \ENDFOR
  \ENDFOR
\end{algorithmic}
\end{algorithm}

Algorithm~\ref{fig:algorithm} shows an entire process of the component search step.
The algorithm starts with a query $Q$ and a file collection $F$.
Since the same file may be included in multiple components, 
we use $F$ to represent a set of existing unique files in a software ecosystem, 
and $Owners(f)$ to represent a set of components including the file $f$.
Our implementation uses SHA-1 file hash to detect files shared by components.

The algorithm computes similarity values $S(q, C)$ between query files and their most similar files in $C$ defined as follows.  
\begin{eqnarray*}
S(q, C) = \max ~ \{ sim(q, f) ~ | ~ f \in C \} 
\end{eqnarray*}
The line 2 initializes $S(q, C)$ to zero and the lines 10 and 11 update it.
A file $f \in C$ may update similarity values of multiple query files, 
because a developer could copy the file $f$ to create the files.


The whole process compares all the pairs of $q \in Q$ and $f \in F$ with two optimizations.
First, it compares the size of trigram sets at the line 5.
The statement uses the following property to avoid unnecessary comparison.
\begin{eqnarray*}
\frac{\min(|X|,|Y|)}{\max(|X|,|Y|)} < th \implies \frac{|X \cap Y|}{|X \cup Y|} < th 
\end{eqnarray*}
The property is derived from $\min(|X|,|Y|) \geq |X \cap Y|$ and $\max(|X|,|Y|) \leq |X \cup Y|$.
Secondly, the process computes $sim_e(q, f)$ that is an estimated similarity computed by $b$-bit minwise hashing technique~\cite{b-bit}.
Since it may have a margin of error, line 6 uses $th-m$ as a threshold, where $m$ specifies allowable errors.  
Finally, the process computes an actual similarity metric $sim(q, f)$ to compare trigrams.
If it is higher than a threshold, components including $f$ are added to $R$ in the component-ranking step.
The lines 10 and 11 record the highest similarity of a query file $q$ and a component $C_i$.
The recorded similarity values $S(q, C_i)$ are used in the component-ranking step.

The algorithm works efficiently because $sim_e(q, f)$ avoids unnecessary actual similarity computation.
In summary, $b$-bit minwise hashing technique approximates a similarity of files by comparing $k$ pairs of $b$-bit signatures.  Each signature represents a trigram sample in a file.  
In case of our implementation, we chose parameters $b=1, k=2048$; 2048 trigram samples in a source file are selected and then translated into 1-bit signatures.  Consequently, a file is represented by a 2048-bit vector.

To compute $sim_e(q, f)$, we use $k$ independent hash functions $h_i(t) ~(1 \leq i \leq k)$.  
Each function translates a trigram $t$ in a file into an integer.  
Our implementation uses 64-bit integers as described in the Appendix.
Using the hash functions, min-hash signatures $m_i(f) ~(1 \leq i \leq k)$ for a file $f$ are computed as follows~\cite{Broder1997}.
\begin{eqnarray*}
m_i(f) = \min ~ \{ h_i(t) ~|~ t \in trigrams(f) \}
\end{eqnarray*}
A min-hash signature $m_i(f)$ represents a trigram sample selected from a file.
If two files $f_1$ and $f_2$ are more similar, more likely $m_i(f_1)$ and $m_i(f_2)$ select the same trigram and result in the same value. 
The probability of $m_i(f_1)=m_i(f_2)$ is represented by the similarity of files~\cite{Broder1997}:
\begin{eqnarray*}
P(\scalebox{0.9}{$\displaystyle m_i(f_1)=m_i(f_2) $}) = sim(f_1, f_2) 
\end{eqnarray*}
We use $b$-bit min-hash signatures $b_i(f) ~(1 \leq i \leq k)$ extracted from min-hash signatures.
In case of $b=1$, the signatures are computed as follows~\cite{b-bit}.
\begin{eqnarray*}
  b_i(f) = LSB( m_i(f) )
\end{eqnarray*}
where $LSB$ is the least significant bit; i.e., $b_i(f) \in \{0, 1\}$.
The probability of $b_i(f_1)=b_i(f_2)$ is represented by
\begin{eqnarray*}
P(\scalebox{0.9}{$\displaystyle b_i(f_1)=b_i(f_2) $}) = sim(f_1, f_2) + \frac{1 - sim(f_1, f_2)}{2} 
\end{eqnarray*}
because the condition is satisfied when $m_i(f_1)=m_i(f_2)$ or two different signatures have the same LSB by chance.

We estimate a similarity of files $q$ and $f$ using their $b$-bit min-hash signatures $b_i(q)$ and $b_i(f) ~(1 \leq i \leq k)$. 
Since the probability of $b_i(q)=b_i(f)$ is dependent on a similarity, 
we compute an estimated similarity $sim_e(q, f)$ from an observed probability as follows.
\begin{eqnarray*}
 sim_e(q, f) & = & (P_{o}(q, f) - \frac{1}{2}) \times 2 \\
 P_o(q, f) & = & 1 - \frac{1}{k} \sum_{i=1}^{k} XOR(b_i(q), b_i(f)) \\
\end{eqnarray*}
where $P_o(q, f)$ is an observed probability of $b_i(q)=b_i(f)$ on $k$ (2048) samples.
The maximum value of $sim_e(f_1, f_2)$ is 1.0.  
If two files are the same, their estimated similarity is always 1.0.
Although the $sim_e(f_1, f_2)$ could be negative, we simply regard them as zero.
It should be noted that this is a simplified version for ease of implementation, compared with the original (strict) estimation that was conducted in~\cite{b-bit}.

\begin{figure}[t]
\centering
\includegraphics[width=\linewidth]{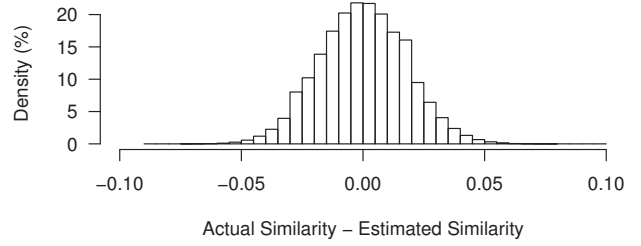}
\caption{The distribution of errors obtained by randomly created samples for $sim(f_1, f_2)=0.6$. }
\label{fig:sim-err-estimated}
\end{figure}

Computation of $sim_e$ is $O(1)$, because it uses $XOR$ bit-operations and bit counting.
It is much efficient than Jaccard index computation that requires $O(n)$ depending on file size $n$.
Since a signature does not change, we compute $b$-bit min-hash signatures for each $f \in F$ and store them in a database.
The tool loads the entire database on memory because it is sufficiently compact; 1 GB memory can store signatures for four million files.
For each query, we then compute signatures for $q \in Q$ and compare them with signatures in the database.
 
The line 6 in Algorithm~\ref{fig:algorithm} uses $th-m$, where $m$ specifies allowable errors.
We use $m=0.1$ for our implementation.  
Fig.~\ref{fig:sim-err-estimated} shows the distribution of errors of $10^8$ randomly created $P_o$ samples under the condition $sim(f_1, f_2) = 0.6$.
In the figure, $sim(f_1, f_2) - sim_e(f_1, f_2)$ is always less than $0.1$.  In other words, $sim_e(f_1, f_2) > th-m$.
We confirmed that it was sufficient in the experiment.

\subsection{Component Ranking}

\begin{table}[t]
\caption{The search result for the example input in Fig.~\ref{fig:concept}}
\label{tab:sim-example}
\centering
\begin{tabular}{l | c | c | c }
\hline
$Q$          & \texttt{X-1.0} & \texttt{X-1.1} & \texttt{Y-0.2} \\ \hline
\texttt{P.c} & 1.0            &  0.7  & 1.0 \\ \hline
\texttt{P.h} & 0.9            &  1.0  & 0.9 \\ \hline
\texttt{Q.c} & 0.9            &  1.0  & 0.9 \\ \hline
\texttt{R.c} & 0.0            &  0.0  & 1.0 \\ \hline
\texttt{S.c} & 0.0            &  0.0  & 0.0 \\ \hline
\hline
$S_Q(C)$   &  2.8 & 2.7  & 3.8   \\ \hline
$|C|$      &  3   & 3    & 5     \\ \hline
\end{tabular}
\end{table}

The second step filters and ranks extracted components $R$ to enable a user to identify an original component easily.
To filter components, we consider that $C_1$ is a better candidate than $C_2$ if $C_1$ provides more similar files than their corresponding source files in $C_2$.
We define this relation $C_1 \supset_{S} C_2$,
\begin{eqnarray*}
C_1 \supset_S C_2  & \iff & (\forall q. S(q, C_1) \geq S(q, C_2) \wedge \\
                   &                 & ~\exists q. S(q, C_1) > S(q, C_2) ~) \vee \\
                   &                 & (\forall q. S(q, C_1) = S(q, C_2)    \wedge |C_1|<|C_2|)
\end{eqnarray*}
where $S(q, C)$ represents similarity values recorded in Algorithm~\ref{fig:algorithm}.
We then select a smaller component (in terms of the number of files) if two components have tied similarity, because it is likely a simpler version.
Using the relation, we select a subset of components $R_S$ from $R$:
\begin{eqnarray*}
R_{S} = \{ C \in R ~ | ~ \nexists C_i \in R. ~ C_i \supset_S C \}
\end{eqnarray*}

Table~\ref{tab:sim-example} shows example similarity values $S(q, C)$ for the example input in Fig.~\ref{fig:concept}.
Since the similarity values satisfy the condition of $\texttt{Y-0.2} \supset_S \texttt{X-1.0}$, 
we obtain $R_S = \{ \texttt{Y-0.2}, \texttt{X-1.1} \}$ excluding \texttt{X-1.0}.

We assign a higher rank to a component that could provide a larger amount of code to the query files.
To measure the degree of potential code reuse, we use the sum of file similarity $S_Q(C)$ defined as follows.
\begin{eqnarray*}
S_Q(C) & = & \sum_{q \in Q} S(q, C) 
\end{eqnarray*}
We rank components in the descending order of $S_Q(C)$.

Our method provides the following information to a user.
\begin{itemize}
\item A list of components $R_S$ sorted by $S_Q(C)$.
Each component is reported with attributes $S_Q(C)$, $|Q|$, and $|C|$.
The result for the example input is following. 
\begin{verbatim}
    Y-0.2  (3.8 / 5)  5 files
    X-1.1  (2.7 / 5)  3 files
\end{verbatim}
A pair of $S_Q(C)$ and $|Q|$ indicates the amount of reused code in $Q$.  
$|C|$ is also important to analyze whether $Q$ is a complete copy of $C$ or not.

\item A full list of components $R$ in the descending order of $S_Q(C)$.
We provide this list because our filtering may accidentally exclude an original component from $R_S$. 
A user can analyze all the components if necessary. 

\item A table of similarity $S(q, C)$.
Although the component search step does not need file names, 
our implementation uses file names for this report.
Table~\ref{tab:sim-table-example} shows an excerpt of a similarity table for \texttt{modules/zlib} directory of \textsl{Firefox 45.0}.
It shows that the analyzed directory is likely a clone of \textsl{zlib 1.2.8} with some modification.
It also shows that file \verb|inflate.c| in Firefox likely includes a similar change as \textsl{MongoDB 3.2.8}.

\end{itemize}
These information enables a user to easily focus on candidates of original components and investigate actual source files in the components.



\begin{table}[t]
\caption{An excerpt of a search result for \texttt{modules/zlib} directory of Firefox 45.0.}
\label{tab:sim-table-example}
\centering
\begin{tabular}{l | l | l }
\hline
$Q$        & zlib 1.2.8   & MongoDB 3.2.8    \\ \hline
gzlib.c    &   1.000 (gzlib.c)    & 0.000    \\ \hline 
inflate.c  &   0.995 (inflate.c)    & 0.999 (third\_party/zlib-1.2.8/inflate.c)   \\ \hline
mozzconf.h &   0.000                & 0.000 \\ \hline 
zconf.h    &   0.985 (zconf.h)    & 0.985 (third\_party/zlib-1.2.8/zconf.h) \\ \hline
zlib.h     &   0.997 (zlib.h)     & 0.997 (third\_party/zlib-1.2.8/zlib.h)  \\ \hline
zutil.c    &   1.000 (zutil.h)      & 1.000 (third\_party/zlib-1.2.8/zutil.h)  \\ \hline
\hline
$S_Q(C)$   &   25.971     & 19.950    \\ \hline
$|C|$      &    51         & 9351    \\ \hline
\end{tabular}
\end{table}

%

\section{Evaluation}
\label{sec:evaluation}

We investigate two research questions to evaluate the effectiveness of our method.
\begin{itemize}
\setlength{\itemindent}{3ex}
\item[\textbf{RQ1.}] Does our method accurately report an original component?
\item[\textbf{RQ2.}] Is our method efficient?
\end{itemize}
To answer the questions, we analyze actual clone-and-own instances in two products: \textsl{Firefox 45.0} and \textsl{Android 4.4.2\_rc1}.
These projects reuse components in a well-organized manner.
Firefox developers often record version numbers of reused components in commit messages in their source code repository.  
Android developers manage their own git repositories for cloned components and record Change-Id to identify changes of the original components.
We manually analyzed directories whose names are likely cloned component names, 
and identified original versions using the commit-log messages. 
We then excluded components whose original versions are unidentifiable.
We spent about one week for the analysis.

Our database of components is the Snapshot Archive of Debian GNU/Linux~\cite{DebianSnapshot}.
We regard a version of a Debian package as a component.
The archive includes all the existing source code packages released for Debian from 2005 until the present.
We automatically downloaded files through its machine usable interface~\cite{DebianSnapshotMUI}.
While Debian package maintainers sometimes apply their own patches, we included only original source tarballs whose names matched a pattern ``\verb|*.orig.*|''.


The database includes 200,018 package files (868 GB in total). 
The resultant dataset includes 9,730,689 C/C++ files and 1,310,235 Java files. 
The total size is 5,733 MLOC (185 GB) including comments and white space.

Our queries comprise 21 directories in Firefox and 54 directories in Android whose original versions are available as Debian packages.
The directory names and corresponding Debian package names are included in the Appendix.
Fig.~\ref{fig:query-size} plots the distributions of the numbers of files in each of directories.
The queries include various size of components; the minimum one comprises two files, while the maximum one comprises 1,163 files.  
The medians of Firefox and Android queries are 76 and 79.5, respectively.
Their total number is 13,720.
Since the database includes copies of Firefox and Android themselves, we exclude their related packages from the search space.

The baseline of evaluation is a simple file search using SHA-1 file hash instead of our similarity function (i.e., $R = \{ C ~ | ~ Q \cap C \neq \phi \}$).  
We sort the extracted components in the descending order of $|Q \cap C|$ that is equivalent to $S_Q(C)$.
Our method uses five threshold for component search: $th=0.6, 0.7, 0.8, 0.9$, and $1.0$ in order to evaluate the effect of threshold.
The parameter $th=1$ ignores white space and comments, differently from the baseline.

For each query, we obtain the rank of the original version in an extracted list.
The rank approximates manual effort of a user.
To evaluate the effect of our filtering method separately from search method, 
we use both our filtering result $R_S$ and a full result without filtering $R$.
In case of $R_S$, we assume that a user investigates all the elements in $R_S$, 
and then investigates $R$ if $R_S$ does not include an original component.


\begin{figure}[t]
\centering
\includegraphics[width=\linewidth]{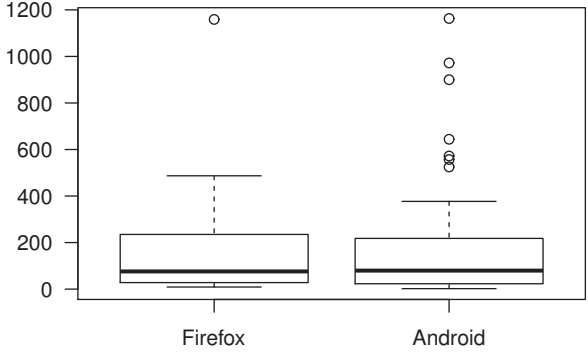}
\caption{The number of files in each query ($|Q|$).}
\label{fig:query-size}
\end{figure}

\subsection{Accuracy}

\begin{figure}[t]
\centering
\includegraphics[width=\linewidth]{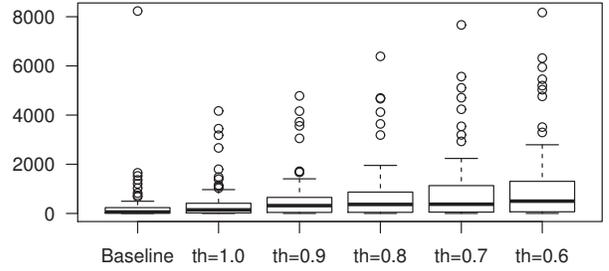}
\caption{The number of components reported for each query ($|R|$).}
\label{fig:raw-count}
\end{figure}

\begin{figure}[t]
\centering
\includegraphics[width=\linewidth]{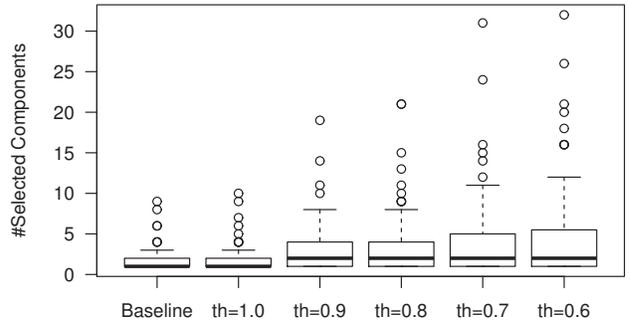}
\caption{The number of selected components for each query ($|R_S|$).}
\label{fig:cluster-count}
\end{figure}


Fig.~\ref{fig:raw-count} shows the distribution of the size of $|R|$, i.e. the number of reported components for each query.   
It shows that many similar files are included in various components.
In case of the baseline method, the median of $|R|$ is 60.
Since our method detects similar files, 
a lower threshold results in a larger number of components.
The medians of $|R|$ are 145, 318, 367, 375, and 500 in cases of $th=1.0$ through $th=0.6$, respectively.  
It should be noted that the plots exclude queries that report no components.  
The baseline method reports no component for three queries.  
One of them is the smallest component whose files are modified.
The other two components have differences in comments and white space.  

Fig.~\ref{fig:cluster-count} plots the number of selected components ($|R_S|$) for each query.
As shown in this figure, the size is less than five components for most of queries.
The median is 1 in cases of the baseline and $th=1.0$.
The median is 2 in cases of $th=0.9$ through $th=0.6$.
While a lower threshold results in a larger $R_S$, the difference is not statistically significant.
Wilcoxon rank sum test results in $p=0.244$ for two cases $th=0.9$ and $th=0.6$.
Our filtering approach successfully selects a small number of components.

\begin{table*}[t]
\caption{Recall of original components in top-$k$ search results.}
\label{tab:recall}
\centering
\begin{tabular}{l  c | c | c | c | c | c | c  }
\hline
Search Method & Ranking & Top-1 & Top-3	& Top-5	& Top-10	& All & Rank Total \\ \hline \hline
Baseline	& $R$  	 	& 0.640	& 0.760	& 0.773	& 0.827	& 0.960	& 931 \\  \hline
$th = 1$	& $R$ 		& 0.653	& 0.773	& 0.787	& 0.853	& 0.987	& 1017 \\ \hline
$th = 0.9$	& $R$   	& 0.680	& 0.800	& 0.827	& 0.867	& 1.000	& 759 \\ \hline
$th = 0.8$	& $R$   	& 0.680	& 0.813	& 0.827	& 0.867	& 1.000	& 818 \\ \hline
$th = 0.7$  & $R$	  	& 0.627	& 0.773	& 0.787	& 0.840	& 1.000	& 930 \\ \hline
$th = 0.6$ 	& $R$		& 0.613	& 0.760	& 0.773	& 0.840	& 1.000	& 928 \\ \hline \hline
Baseline    & $R_S$		& 0.707	& 0.840	& 0.840	& 0.867	& 0.960	& 719 \\ \hline
$th = 1$ & $R_S$		& 0.733	& 0.867	& 0.867	& 0.893	& 0.987	& 785 \\ \hline
$th = 0.9$ & $R_S$		& \textbf{0.733} & \textbf{0.893} & \textbf{0.907}	& \textbf{0.920}	& \textbf{1.000}		& \textbf{551} \\ \hline
$th = 0.8$ & $R_S$		& 0.733	& 0.880	& 0.893	& 0.920	& 1.000	& 627 \\ \hline
$th = 0.7$ & $R_S$		& 0.680	& 0.853	& 0.880	& 0.907	& 1.000	& 692 \\ \hline
$th = 0.6$ & $R_S$		& 0.667	& 0.853	& 0.880	& 0.907	& 1.000	& 689 \\ \hline \hline
\end{tabular}
\end{table*}

Table~\ref{tab:recall} summarizes recall of components appeared in the top-$k$ elements of resultant lists.
For example, the baseline method ranks 64\% of original components at the top of lists.
The column ``All'' shows recall of components in the entire lists.
The column ``Rank Total'' shows the sum of positions of the original components in the results. 
It approximates the effort to identify all the original components in the lists.
Each row shows the result of a configuration.  
Each configuration uses a search method in the first column.
The top six configurations simply reports a full list of $R$ without our filtering.
The bottom six configurations use a filtered list of $R_S$.
The result shows that our method with $th=0.9$, $R_S$ performs the best among the configurations.
A user can identify all the original components in the top-5 components for 90.7\% cases and 551 components in total. 
Our method reduces 40\% of user's effort compared with the baseline reporting 931 components.
The baseline method also requires additional effort to analyze three queries that resulted in no components.

The result shows that modified files are important to identify original components.
A simple exclusion of white space and comments does not improve the result; 
our similar file search with $th=1$ and $R$ results in a longer list of components, although it slightly improves recall.
Our method with $th < 1$ performs better than the baseline and $th=1$, 
because modified files provide a clue to sort components and identify an original version.
A lower threshold does not improve a result, because it also detects various versions of the same components affecting the results.

Our filtering method successfully improves results in all the cases.
In case of $th=0.9$, it reduces 27\% of manual effort to investigate the reported lists.
This is because the most similar version of a library among several versions in $R$ is actually the original version.
In this experiment, only 882 out of 13720 query files (6.4\%) are modified in Firefox and Android.
Developers tend to reuse source files without modification.

\begin{table}[t]
\caption{An excerpt of a search result of libpng component in Android~4.4.2\_rc1 with $th=0.9$.}
\label{tab:libpng}
\begin{center}
\begin{tabular}{l l l}
\hline
Query files	& libpng 1.2.46 & libpng 1.2.49 \\ \hline
png.c		& 0.962 & 0.958 \\ \hline
png.h		& 0.982 & 0.980 \\ \hline
pngpread.c	& 1	& 0 \\ \hline 
pngrtran.c	& 1	& 0.999 \\ \hline
pngrutil.c	& 0.985 & 0.978 \\ \hline
pngset.c	& 0.987 & 1.000 \\ \hline
pngtest.c	& 1	& 0.999 \\ \hline
\hline
$S_Q(C)$	& 41.902 & 39.900 \\ \hline
$|C|$		& 45	& 45 \\ \hline
\end{tabular}
\end{center}
\end{table}

We do not regard other components in $R_S$ as false positives, because some of them are also informative.
For example, Android includes a modified version of \textsl{libpng 1.2.46}.  
Our method ranks the original version at the top.
Our method also ranks \textsl{libpng 1.2.49} at the second.
Table~\ref{tab:libpng} shows an excerpt from the reported similarity values for them.
It shows that file \verb|pngset.c| in Android is the same as a file in \textsl{libpng 1.2.49}.
The file includes a security fix for CVE-2011-3048~\cite{CVE-2011-3048} found in \textsl{1.2.x} before \textsl{1.2.49}.
Our method successfully reports that the cloned component includes a part of the newer version.


Our filtering method excluded some original components from $R_S$.
An example is the Expat XML parser library in Android. 
Our method reports \textsl{node-expat-2.3.12} that is a NodeJS binding component instead of the original version,
because both Android and the component includes all the files in the original Expat library and an additional header file \verb|expat_config.h| generated by \verb|configure| script.
Another example is \textsl{zlib 1.2.8} in Android.
In this case, \textsl{zlib 1.2.8} component in the database does not include \verb|contrib| files because of a license issue.  Hence, our method accidentally reports another component including a full copy of the original version.
This is a limitation of our repository-based approach.

Our file search cannot identify any origins for 1240 files (9.04\%) using $th=0.6$.
They are likely added by the projects.
We believe that this is also useful for a user to understand how a cloned component is modified.
The baseline method does not provide this additional information.


\subsection{Performance}
\label{sec:performance}

\begin{figure}[t]
\centering
\includegraphics[width=\linewidth]{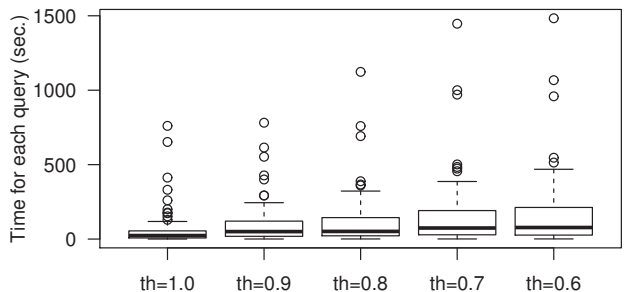}
\caption{Time spent for each query.}
\label{fig:time-per-query}
\end{figure}

\begin{table}[t]
\caption{Time for 75 queries and the numbers of computed similarity values.}
\label{tab:time}
\begin{center}
\begin{tabular}{l | r | r | r | r}
\hline
            & \multicolumn{2}{|c|}{Time (sec.)} & & \\ \cline{2-3}
Method		& Median & Total  & \#$sim_e(q, f)$ & \#$sim(q, f)$ \\ \hline
Baseline 	& 2.0 & 270	&	N/A	& N/A \\ \hline 
$th=1$		& 23.7 & 5,013	&   93,836,051 & 27,300  \\ \hline 
$th=0.9$	& 50.1 &	7,719	&	5,468,450,021 & 72,848  \\ \hline 
$th=0.8$	& 51.4 &	8,956	&	11,463,856,656 & 98,512  \\ \hline 
$th=0.7$	& 74.3 &	11,951	&	18,229,604,035  & 125,520  \\ \hline 
$th=0.6$	& 77.7 & 	12,595	&	25,967,936,034 & 189,043  \\ \hline 


\end{tabular}
\end{center}
\end{table}

Our experiment is performed on a workstation equipped with 
Intel Xeon E5-2690 v3 (2.6 GHz), 64 GB RAM, and 2 TB HDD.
We use a single thread to execute a query.

Figure~\ref{fig:time-per-query} shows the time spent for each query with different threshold.
A lower threshold takes longer time, because it affects a size-based optimization of Algorithm~\ref{fig:algorithm} (Line 5).
The median is 77.7 seconds for $th=0.6$.  The longest query takes 25 minutes.  
Table~\ref{tab:time} shows the median time required for each query and the total time for 75 queries for each parameter.
The baseline method takes about 200 seconds to read files and compute SHA-1 file hash values.
Comparison of SHA-1 file hash takes 70 seconds.
While our method takes significantly longer time than the baseline, 
the time is still practical because a user can analyze a result of a query during execution of other queries.
It should be noted that the query time is strongly dependent on component search.
The component-ranking step takes less than 2 seconds in all the configurations.  

\begin{figure}[t]
\centering
\includegraphics[width=\linewidth]{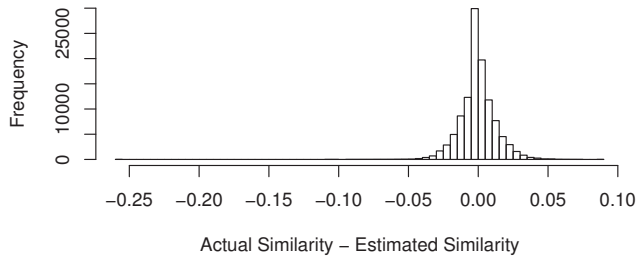}
\caption{The distribution of errors for file pairs compared during the experiment with $th=0.6$.}
\label{fig:sim-err}
\end{figure}

Our method compares each query file with a database of files.
Since the total number of query files is 13,720, 
our method takes at most a few seconds to compare a file with 10 million files.
This performance is achieved by $b$-bit minwise hashing technique.
To analyze the effect of similarity estimation, we count the numbers of computed estimated and actual similarity values.  They are shown in the columns \#$sim_e(q, f)$ and \#$sim(q, f)$ of Table~\ref{tab:time}.
The table clearly shows that $sim_e(q, f)$ enables us to avoid the computation of $sim(q, f)$.
The ratio of \#$sim(q, f)$ is reported to be less than 0.03\% in all cases.

The experiment uses $m=0.1$ in Algorithm~\ref{fig:algorithm}. 
We executed the same experiment with $m=0.2$ and confirmed the same result.  
Hence, $m=0.1$ is sufficiently large.
Fig.~\ref{fig:sim-err} shows the distribution of errors between actual and estimated similarity.
The lowest estimated similarity satisfying the $sim(q, f) \geq 0.6$ is $sim_e(q, f) = 0.513$.

Our implementation constructs a database of components in prior to search.
We took about 5 days to extract all the files from package archives, remove duplicated files, and compute file signatures.
The archive extraction step is the bottleneck, because the step has to process 2 TB of files in archives.
Our database uses 4 GB for file signatures and 20 GB for component names and file names.
Our implementation keeps file signatures on memory.
Our database can be incrementally updated, by simply adding file signatures.
A larger database can be hosted by multiple servers, 
because we can search multiple component sets independently and later merge the final result.

\section{Threats to Validity}
\label{sec:threats}

We analyzed Firefox and Android source code repositories,
because developers in the projects keep their record of code reuse.
Our result might be dependent on code reuse strategies of the projects.

We manually analyzed commit messages recorded in the repositories.
Since the version information is verified by two authors but no developers of the projects, 
the result has a risk of human error.



The result is dependent on components included in a database.
Our database represents a single software ecosystem: Debian GNU/Linux source code packages.
Since we used original source code provided by each project, we believe the packages reflect activities of various projects.
On the other hand, our analysis does not reflect source code modified by package maintainers of the operating system.

The collection of software packages may miss important packages.
For example, \textsl{libpng} project maintains a number of branches: \textsl{libpng 0.x, 1.0.x, 1.2.x, 1.4.x, 1.5.x, 1.6.x,} and \textsl{1.7.x}.
Since package maintainers selected major branches to create Debian packages, 
our dataset includes a subset of official versions. 
It may affect the analysis of variants in the experiment.

While the Debian Snapshot Archive is publicly available, 
we found several errors (i.e., 404 not found) and corrupted archive files during our analysis.
Since the entire dataset is large, the result might be affected by this accidental data corruption.

The performance of $b$-bit minwise hashing signature is dependent on underlying hash functions.
It may miss a similar file with a very low probability.
We confirmed that we did not miss any files during the experiment.
For replicability, our strategy to define 2048 hash functions is included in Appendix.

\section{Conclusion}
\label{sec:conclusion}

This paper proposed a code search method to extract original components from a software ecosystem.
In the experiment, our method successfully reported original components compared with the baseline method.
Our implementation also reports the computed similarity values to enable further analysis.

To implement an efficient code search, we used $b$-bit minwise hashing technique.
It enabled us to extract less than 0.03\% of likely similar files from a database in a second.
Our method also introduced a component filtering method using aggregated file similarity.
It reduces manual effort to analyze reported components.

In future work, we would like to apply our method to analyze clone-and-own reuse activity in various projects including industrial organizations.
We are also interested in a systematic method to analyze known issues and vulnerabilities caused by cloned components in a software product.


\section*{Acknowledgment}

This work was supported by JSPS KAKENHI Grant Numbers JP25220003, JP26280021, and JP15H02683.

We are grateful to Naohiro Kawamitsu for the implementation of hash functions and Raula Gaikovina Kula for his comments to improve the manuscript.


\appendix

\textbf{Hash Function. }
Our implementation uses 2048 hash functions to translate a file into a file signature.
We use the following hash functions on 64-bit integer:
\begin{eqnarray*}
h_i(t) = a_i \times base(t) + b_i 
\end{eqnarray*}
where $1 \leq i \leq 2048$, $a_i$ and $b_i$ are randomly generated 64-bit integers.
The $base(t)$ function translates a trigram into a 64-bit integer. 
Since a trigram in a multiset is identified by four elements \texttt{A}, \texttt{B}, \texttt{C}, and $i$ ($i$-th occurrence of a trigram \texttt{ABC}), the $base$ function is defined as follows.
\begin{eqnarray*}
base(\texttt{A}, \texttt{B}, \texttt{C}, i) = & (((i \times 65537) + \texttt{A.hashCode()}) \\
     & \times 65537 + \texttt{B.hashCode()})  \\
     & \times 65537 + \texttt{C.hashCode()}   
\end{eqnarray*}
where \verb|hashCode()| is \verb|String.hashCode| method in Java.


\textbf{Component List.  }
Table~\ref{tab:component-list} and Table~\ref{tab:component-list-2} show analyzed directories in Firefox and Android, respectively.
The package names are defined by Debian GNU/Linux package maintainers; some of them are different from official project names.


\begin{table}[t]
\begin{center}
\caption{Analyzed directories in Firefox 45.0}
\label{tab:component-list}
\begin{tabular}{l l}
\hline
Package Name & Directory \\ \hline \hline
cairo-1.10 & gfx/cairo \\ \hline
double-conversion-1.1.1 & mfbt/double-conversion \\ \hline
graphite2-1.3.6 & gfx/graphite2 \\ \hline
gtest-1.6.0 & testing/gtest  \\ \hline
hunspell-1.3.3 & extensions/spellcheck/hunspell \\ \hline
libav-11.3 & media/libav \\ \hline
libevent-2.0.21 & ipc/chromium/src/third\_party/libevent  \\ \hline
libffi-3.1 & js/src/ctypes/libffi \\ \hline
libjpeg-turbo1.4.2 & media/libjpeg \\ \hline
libpng1.6-1.6.19 & media/libpng \\ \hline
libsoundtouch-1.9.0 & media/libsoundtouch, \\ \hline
libspeex-1.2 & media/libspeex\_resampler \\ \hline
libvorbis-1.3.5 & media/libvorbis \\ \hline 
libvpx-1.4.0 & media/libvpx \\ \hline
nspr-4.12, & nsprpub \\ \hline
nss-3.21.1 & security/nss \\ \hline
opus-1.1 & media/libopus \\ \hline
snappy-1.0.4 & other-licenses/snappy \\ \hline
srtp-1.4.4 & netwerk/srtp\\ \hline
stlport-5.2.1 & build/stlport \\ \hline
zlib-1.2.8 &  modules/zlib \\ \hline
\end{tabular}
\end{center}
\end{table}

\begin{table}[t]
\begin{center}
\caption{Analyzed directories in Android 4.4.2\_rc1}
\label{tab:component-list-2}
\begin{tabular}{l l | l l}
\hline
Package Name & Directory$\dagger$ & Package Name & Directory$\dagger$  \\
\hline \hline
arduino-0022 & arduino & libogg-1.2.0 & libogg \\ \hline
blktrace-1.0.1 & blktrace & libpng-1.2.46 & libpng \\ \hline
bouncycastle-1.49 & bouncycastle & libsepol-2.2 & libsepol \\ \hline
bsdiff-4.3 & bsdiff & libusb-1.0.8 & libusb \\ \hline
bzip2-1.0.6 & bzip2 & libvorbis-1.3.1 & libvorbis \\ \hline
checkpolicy-2.1.11 & checkpolicy & libvpx-1.3.0 & libvpx \\ \hline
dnsmasq-2.51 & dnsmasq & libxml2-2.7.8 & libxml2 \\ \hline
dropbear-0.49 & dropbear & libxslt-1.1.26 & libxslt \\ \hline
e2fsprogs-1.41.14 & e2fsprogs & mksh-43 & mksh \\ \hline
easymock-2.5.2 & easymock & netperf-2.4.4 & netperf \\ \hline
expat-2.1.0 &  expat & openfst-1.3.3 & openfst \\ \hline
flac-1.2.1 & flac & openssh-5.9 & openssh \\ \hline
genext2fs-1.4.1 & genext2fs & openssl-1.0.1e & openssh  \\ \hline
grub-0.97 & grub & oprofile-0.9.6 & oprofile \\ \hline
guava-libraries- & guava & pixman-0.30.0 & pixman \\ 
11.0.2 6 & & &  \\ \hline
harfbuzz-0.9.14 & harfbuzz\_ng & ppp-2.4.5 & ppp \\ \hline
ipsec-tools-0.7.3 & ipsec-tools & protobuf-2.3.0  & protobuf \\\hline
iptables-1.4.11.1 & iptables & safe-iop-0.3.1 & safe-iop \\\hline
iputils-3:20121221 & iputils & scrypt-1.1.6 & scrypt \\\hline
jhead-2.86 & jhead & speex-1.2rc1 & speex \\\hline
jpeg-6b & jpeg & srtp-1.4.4 & srtp \\\hline
jsilver-1.0.0 & jsilver & stlport-5.2.1 & stlport \\\hline
junit4-4.10 & junit & stressapptest- & stressapptest \\
            &       & 1.0.4 &  \\ \hline
libgsm-1.0.13 & libgsm & tagsoup-1.2 & tagsoup \\ \hline
libhamcrest-java- & hamcrest & tcpdump-3.9.8 & tcpdump \\ 
 1.1 & & & \\ \hline
libmtp-1.0.1 & libmtp & valgrind-3.8.1 & calgrind \\ \hline
libnl-2.0 & libnl-headers & zlib-1.2.8 & zlib \\ 
\hline
\multicolumn{4}{l}{$\dagger$ All the component directories are located in \texttt{external} directory.} \\
\end{tabular}
\end{center}
\end{table}

\clearpage

\newpage 


\end{document}